# Low-Terahertz Channel Performance in Falling Wheat for Moisture Sensing

Xiaoxiang Li, Hanchen Liu, Kefeng Huang, Jiabiao Zhao, Mingxia Zhang, Jie Yang, Yuheng Song, Wenbo Liu, Peian Li, Chunhua Zhu, *Member, IEEE,* Jianjun Ma, *Member, IEEE*

***Abstract*—Terahertz (THz) waves are highly sensitive to water and therefore offer a promising noncontact approach to wheat moisture sensing, but existing THz studies have been limited to static samples. This article investigates moisture sensing based on the performance of a THz channel in falling wheat, as encountered at grain intake. The channel power loss is measured at six frequencies from 120 to 320 GHz for three moisture contents and interpreted with a physics-based model that couples the double-Debye permittivity of water, Landau-Lifshitz-Looyenga mixing, and Lorenz-Mie extinction. The model reproduces the measured loss. Above the extinction maximum, the dielectric moisture response peaks where the wavelength approaches the kernel diameter, and that a two-frequency loss ratio removes the loading dependence. A channel-level analysis of QPSK and 16-QAM links shows that a 1 dB moisture-induced loss change alters the bit error rate (BER) by a factor of 3 to 30. THz communication channels at grain intake could therefore sense moisture as a byproduct of data transmission.**

## I. INTRODUCTION

Wheat moisture content (MC) is a critical factor influencing storage stability and end-use quality. Excess moisture shortens the safe storage time and degrades protein quality [1, 2], and wheat is generally kept within about 8%-14% MC [3]. Moisture measurement techniques comprise conventional reference methods and nondestructive sensing methods [4, 5]. Gravimetric air-oven methods provide the reference values [6, 7] but are destructive and time consuming, and Karl Fischer titration requires chemical reagents, calibration, and specialized operation [8]. Contact-based electrical meters are faster, but their readings depend on wheat type, temperature, bulk density, and operating frequency [9, 10].

Free-space microwave systems, metasurface antennas, ultra-wideband radar, and millimeter-wave radar have been developed for nondestructive or noncontact measurement of wheat moisture [11-16], and near-infrared and hyperspectral techniques provide moisture-related absorption information [17, 18]. A central lesson of the microwave literature is that the transmitted signal depends on the bulk density, or loading, as well as on moisture, which led to density-independent calibration functions combining two measured quantities [19, 20]. With them, the moisture of flowing grain has been determined to within about 1% MC, although the accuracy degrades in dilute flow [21, 22]. The terahertz (THz) band offers a complementary modality. Its short wavelength enables fine spatial resolution with compact, highly directive apertures [23, 24], and THz radiation is intrinsically sensitive to water because of the strong response of its hydrogen-bond network [25, 26]. Fixed-frequency sensors near 200 GHz already resolve moisture in paper production from the transmission loss alone [27], and THz spectroscopy is an established laboratory tool for moisture in agri-food products [28]. In crops, THz studies have progressed from leaf water status [29, 30] to wheat grain, where THz time-domain spectroscopy (TDS) has been used to characterize crushed wheat samples [31, 32] and deep-learning-assisted THz imaging visualizes the water distribution within kernels [33, 34].

These studies, however, rely on static samples and prescribed geometries, whereas wheat at intake and transfer is in motion and, before bulk accumulation, in a sparse falling state. Falling wheat offers a geometrically bounded path without sampling, as free-space microwave studies of flowing grain have shown [14, 35]. At low-THz frequencies, which span roughly 100-300 GHz, it also poses an unaddressed propagation problem. The wavelength is comparable to the kernel size, so that each kernel is a Mie scatterer rather than a small absorbing inclusion, and scattering, not only water absorption, governs the loss [36, 37]. The situation resembles THz channel propagation through rain, snow, and other particle-laden media [38, 39], where Lorenz-Mie extinction integrated over the particle population describes the measured attenuation [40-42]. Millimeter wave radar has also been applied to free falling particle streams [43]. To the best of our knowledge, however, no THz channel measurement or model exists for falling grain, and the bit error rate (BER) of a THz channel has not been examined as a reliable moisture observable.

This article addresses this gap by investigating moisture sensing through the channel power and BER performance of a low-THz channel across falling wheat, combining channel measurements with physics-based modeling. Section II describes the measurement system and the dielectric and extinction model and compares the predicted and measured channel power loss. Section III analyzes the moisture sensitivity of the channel power by simulation, derives frequency-selection and loading-independent sensing rules, and characterizes the measured received-power fluctuations. Section IV evaluates the BER performance and its use as a moisture indicator, and Section V concludes the article.

## II. EXPERIMENTAL SETUP AND THEORETICAL MODELING

As shown in Fig. 1(a), the measurements used a fixed, highly directional line of sight (LoS) channel whose lens-equipped horn antennas concentrate the fields in the falling wheat produced by a wheat feeder, emulating the flow during loading into a storage facility. The total channel distance between the transmitting and receiving antennas was 2.35 m. The falling wheat was located 1.32 m from the transmitter and 1.03 m from the receiver along the beam axis. The effective width of the falling-wheat stream used for the areal density calculation was 0.11 m. The drop height before the wheat reached the beam center was 28.8 cm, and the total length of the grain curtain was 62 cm. The beam diameter at the wheat stream, determined by the lens aperture, was 11.07 cm. A Ceyear 1465D signal generator, which operates from 100 kHz to 20 GHz [Fig. 1(c)], drove one of two multiplier chains [Fig. 1(d)] - a Ceyear 82406B (×12) with an HD-1400SGAH25 horn for 110-170 GHz, and a Ceyear 82406D (×18) with a Ceyear 89901S horn for 220-325 GHz. Each horn was followed by a dielectric lens (high-density polyethylene in the upper band). The horn-lens antenna had a gain of approximately 33 dBi at 220 GHz and a beamwidth of approximately 5.7°. The received power was detected by a Ceyear 71718 power sensor [Fig. 1(e)] and read by the power meter [Fig. 1(b)] at a rate of 5 Hz. The instruments were controlled by a host computer through LAN-SCPI commands, with custom instructions over an RS-485 bus coordinating control and acquisition, and all data were recorded continuously to capture the wheat-induced power fluctuations. The transmitter and receiver were 88 cm above the ground, far beyond the first Fresnel-zone radius at all operating frequencies.

Each measurement record consisted of three segments: free-space transmission during the first 6 s, transmission through the falling wheat, and free-space transmission again after the flow had stopped. Only the stable falling interval of the second segment was analyzed, and the power loss was calculated as the difference between the free space received power and the received power through the falling wheat at the same frequency, so that the antenna gains cancel. At every frequency, the received power returned to its initial free space level after the flow stopped, which confirms that the system did not drift within a record. At 220 GHz, a 25 s free space record showed a standard deviation of 0.005 dB and a peak-to-peak variation of 0.025 dB around a mean of -10.47 dBm.

The wheat moisture contents (MC) of 7.8%, 11%, and 14% were prepared by conditioning wheat samples to target moisture contents before the falling wheat experiments. Wheat samples were wetted with water and stored in sealed containers for 72 h to reach moisture equilibrium. The three moisture levels were measured on a wet basis using an LDS-1G capacitance-type grain moisture meter. For each moisture condition, wheat samples were taken from the conditioned batch, and the moisture content was measured repeatedly, with the reported value being the average of repeated measurements. Because the amount of wheat delivered at a fixed feeder setting varies with MC, the loading was determined for each sample. The relation between wheat mass and kernel count was fitted at each MC, and the mass delivered within 10 s then gave 17063, 16134, and 20339 kernels for MC of 7.8%, 11%, and 14%, respectively.

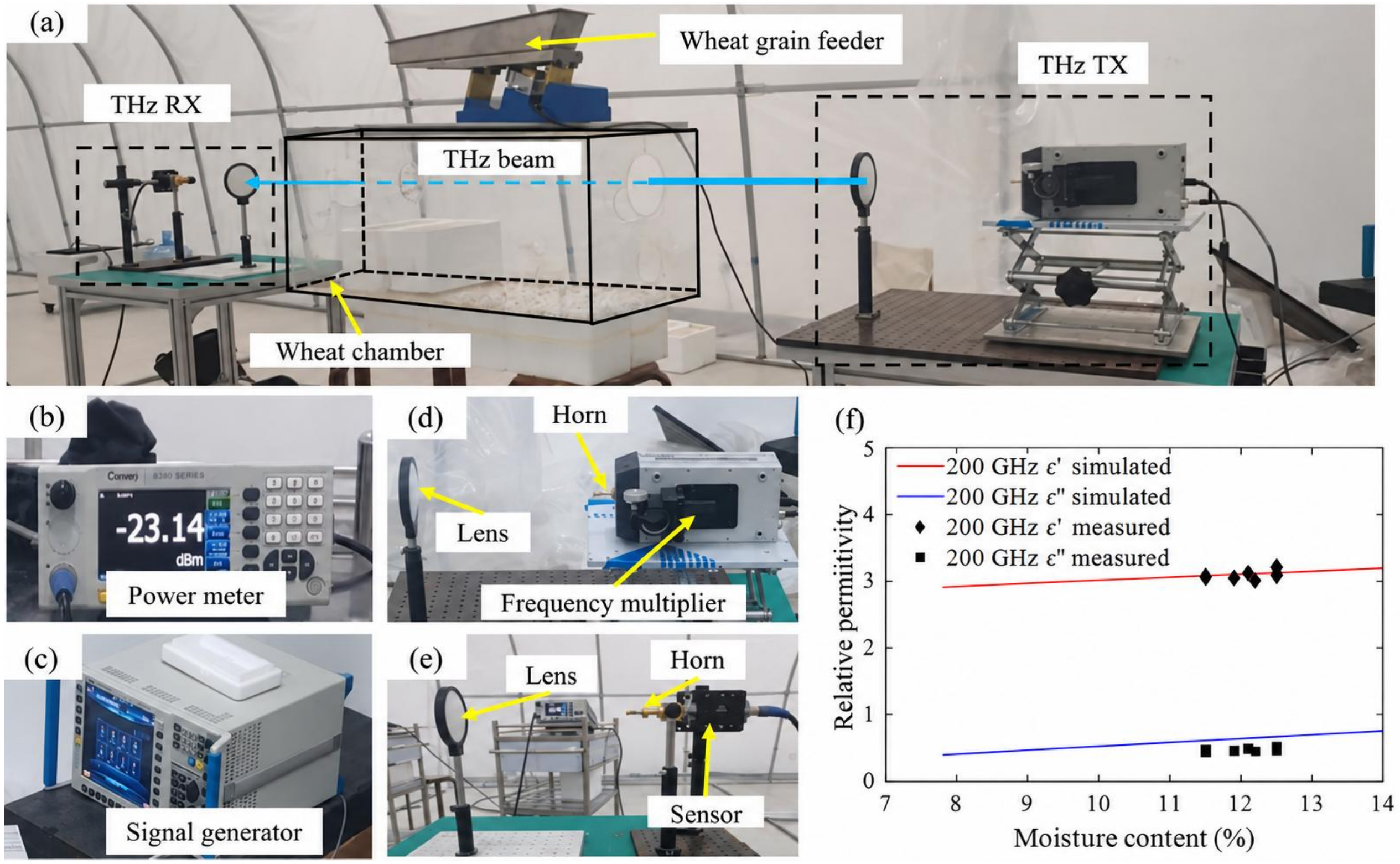

Fig. 1. (a) Experimental setup for channel measurement through falling wheat with (b) power meter, (c) signal generator, (d) THz transmitter side, (e) THz receiver side, and (f) simulated permittivity of wheat kernels versus moisture content at 200 GHz, compared with values digitized from [51].

Power loss in falling wheat is governed by the moisture-dependent dielectric response of the kernels and the collective extinction of the kernel population. Following the standard approach to THz moisture quantification [44], the complex permittivity of water is described by the double-Debye model [45], as

$$\varepsilon_w(f,T) = \varepsilon_2 + \frac{\varepsilon_0 - \varepsilon_1}{1 - if/f_1} + \frac{\varepsilon_1 - \varepsilon_2}{1 - if/f_2} \quad (1)$$

where $f$ denotes the frequency, $T$ denotes the temperature, $\varepsilon_0$, $\varepsilon_1$, and $\varepsilon_2$ are the static, intermediate, and high-frequency permittivities, and $f_1 = 1/(2\pi\tau_1)$ and $f_2 = 1/(2\pi\tau_2)$ are the relaxation frequencies, with the relaxation times $\tau_1$ and $\tau_2$. In all models, the water permittivity was evaluated with a fixed room temperature parameter set corresponding approximately to $T$ = 25 °C, close to the room temperature during the measurements: $\varepsilon_0 = 78.3$, $\varepsilon_1 = 5.2$, $\varepsilon_2 = 3.3$, $\tau_1 = 8.3$ ps, and $\tau_2 = 0.18$ ps [45]. Moist wheat is treated as a binary mixture of a dry wheat matrix and water, with an effective permittivity given by the Landau-Lifshitz-Looyenga (LLL) model [46, 47], a mixing rule suited to irregular inclusions that has been tested on air-particle mixtures of granular wheat and flour [48] and applied to moist flour matrices [49],

$$\varepsilon_{\text{eff}}(f,M) = \left[ v_d \varepsilon_d^{1/3} + v_w \varepsilon_w^{1/3}(f,T) \right]^3 \quad (2)$$

where $v_d$ and $v_w$ are the volume fractions of the dry matrix and water, obtained from the wet-basis MC, $M$, as $v_w = (M/\rho_w)/[M/\rho_w+(1-M)/\rho_d]$ and $v_d = 1-v_w$, with $\rho_w$ =997 kg/m$^3$ and $\rho_d$ =1350 kg/m$^3$, an effective dry-matrix density taken from single-kernel measurements [50]. The dry matrix permittivity $\varepsilon_d = 2.54$ was fitted to the permittivity of eight wheat varieties at 11.5%-12.5% MC, derived from their refractive index and absorption coefficient [51]. The model then reproduces the real part [Fig. 1(*f*)] but overestimates the imaginary part, even for a lossless matrix. So Im($\varepsilon_d$) was set to zero. This discrepancy affects the predicted attenuation little but the magnitude of the moisture sensitivity more (see Section III-B). Because these data span only about one percentage point of MC, they fix $\varepsilon_d$ but not the moisture dependence, which follows from Eq.(1). Its parameters characterize pure water independently of the present data over the entire frequency range, and its temperature dependence [45] allows thermal effects to be evaluated (see Section III-C). Combined with Eq. (2), the effective permittivity thus depends on moisture only through $v_w$, without additional fitting parameters. The largely bound state of water at 8%-14% MC, which responds more weakly than free water [10, 52], mainly affects the damping of the extinction features, governed by the absorption index, and less their positions, governed by the refractive index, which the dry matrix dominates.

The kernel geometry follows the three-axis description of [53]. For the variety Zhongmai 175 the mean length, width, and thickness are approximately 6.61, 3.56, and 3.108 mm, and the thickness follows a Gaussian distribution, N(3.108, 0.156$^2$) mm, in line with the approximately Gaussian statistics of kernel traits reported in [54]. The kernels are represented by homogeneous spheres whose radius distribution is parameterized from the thickness distribution, with a mean radius of 1.554 mm. The reference diameter $D_c$ = 3.108 mm was taken at 8% MC. Moisture-induced swelling was also included by allowing the effective diameter to increase with MC, using an approximate projected-area expansion rate of 0.078 mm$^2$ per 1% MC estimated from the linear projected-area trend reported in [55]. This corresponds to a diameter increase of approximately 0.016 mm per 1% MC, i.e., about 0.5% of the diameter and about 1% of the geometric cross section per 1% MC. Although small compared with the run-to-run differences in loading, this geometric change is comparable to the dielectric moisture response and, at higher frequencies where the kernels become opaque, dominates the predicted moisture sensitivity. Predictions in that regime therefore depend directly on the assumed swelling law. Lorenz-Mie theory then provides the extinction for any size parameter [56-58], and the specific attenuation of the falling wheat, which is expressed in decibels per unit length, follows by integration over the size distribution [59, 60].

$$\alpha = 4.343 \int_{r_{\min}}^{r_{\max}} N(r)\,\sigma(r)\,\mathrm{d}r \quad (3)$$

where $N(r)$ is the number-density distribution of kernels with radius $r$ and $\sigma(r)$ is the extinction cross section of an individual equivalent sphere.

$$\sigma(r) = \pi r^2 Q(r) \quad (4)$$

$$Q(r) = \frac{2}{x^2} \sum_{n=1}^{\infty} (2n+1)\,\mathrm{Re}(a_n + b_n) \quad (5)$$

where $Q$ is the extinction efficiency, $x = 2\pi r/\lambda$ is the size parameter, $\lambda$ is the wavelength, and $a_n$ and $b_n$ are the Mie coefficients. The extinction efficiency $Q$ accounts for both absorption and scattering.

Equation (3) describes coherent attenuation by independent scatterers under three assumptions. First, the medium is tenuous. Corrections grow with volume fraction and optical depth $\tau = \alpha L_{fg}/4.343$, where $L_{fg}$ is the channel length in falling wheat [60, 61]. In our measurements, $\tau \approx 0.4$-$0.7$. Second, no forward-scattered power is collected. By Lorenz-Mie integration over the antenna main lobe, the forward collection effect reduces the apparent extinction by at most 2% at 120 GHz and 10%-14% at 320 GHz and is treated as a model uncertainty [58, 62]. Third, the elongated kernels [63] are modeled as thickness-based effective spheres, as is common for hydrometeors and granular media [37, 64-66] The model thus targets the frequency dependence and relative moisture trend, and shape effects enter the uncertainty of the absolute loss, which is tied to the calibrated areal density.

Fig. 2 compares the measured power loss with the values predicted by the model, in which the kernel count of each sample enters as the loading input. For a mean kernel radius of 1.554 mm, the six measurement frequencies correspond to $x = 2\pi rf/c \approx 3.9$-$10.4$, with the kernel diameter exceeding the wavelength throughout this range. For all three moisture levels, the mean measured loss of about 1.8-2.9 dB follows the prediction with a root-mean-square (RMS) deviation of about 0.15 dB over the 18 frequency-moisture combinations (at most about 0.3 dB). The remaining deviations can be attributed to the dynamic spatial distribution of the falling wheat, variations in the instantaneous kernel number density, the spherical approximation, and the forward-collection effect discussed above. Following the practice for rain and snow, in which Lorenz-Mie predictions based on the particle loading and size statistics are compared with the measured attenuation [39-42], the agreement indicates that the dielectric-mixing and extinction model provides a reasonable first-order description of THz power loss in falling wheat, which will be employed in the following sections.

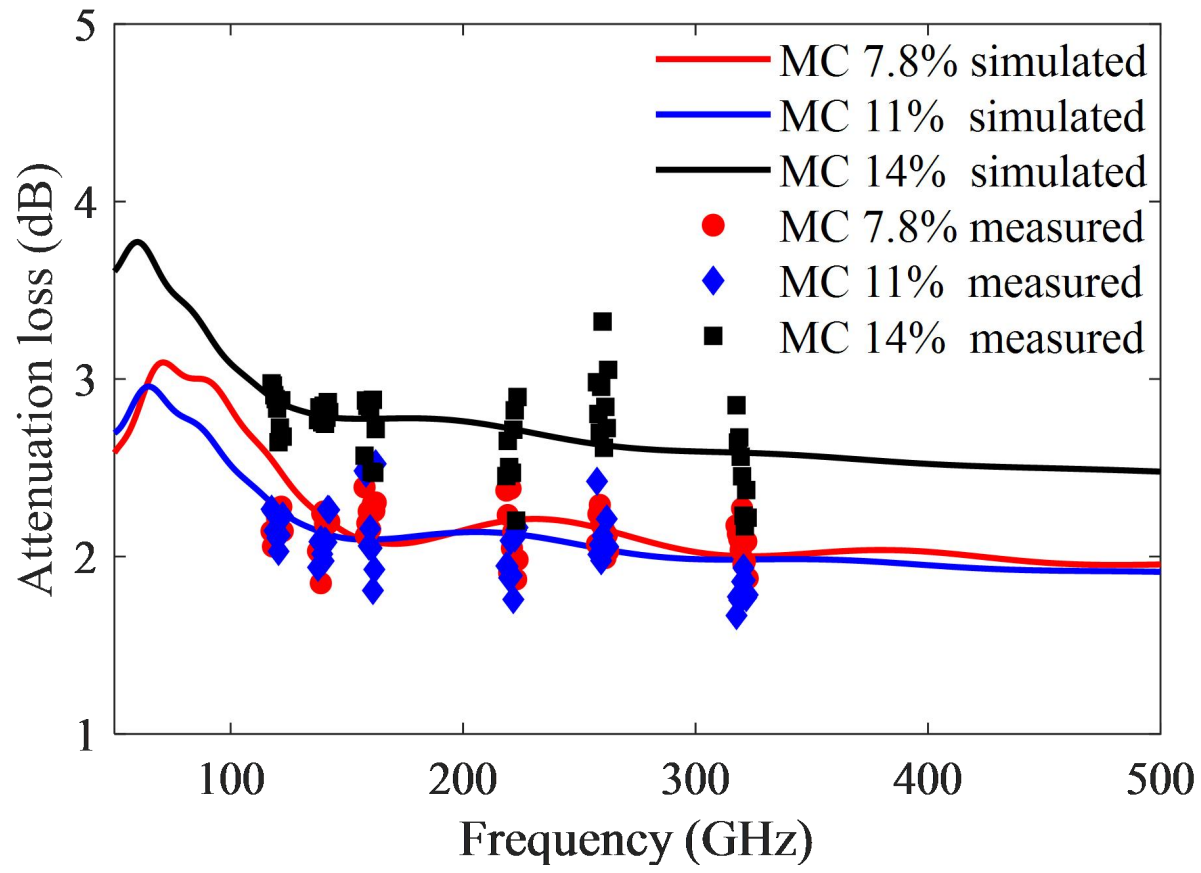

Fig. 2. Measured and predicted power loss through falling wheat versus frequency for different moisture contents. The predictions use the measured kernel counts per 10 s of 17063, 16134, and 20339 for MC of 7.8%, 11%, and 14%.

## III. Moisture Sensitivity of the Falling-Wheat Channel

### A. Loss Spectra

Fig. 3 shows the power loss predicted by the model of Section II and its moisture sensitivity from 50 to 400 GHz for MC of 8%-14%, the safe-storage range [3], a 15-cm channel in falling wheat, a kernel number density of $N = 1.6\times10^6$ m$^{-3}$ (an areal density of $2.4\times10^5$ m$^{-2}$), and characteristic diameters $D_c$ of 2.55, 3.108, and 4.25 mm with the Gaussian size distribution again. These diameters span the effective equivalent sphere sizes of wheat kernels, from below the kernel thickness to the volume-equivalent diameter of the mean kernel, about 4.2 mm [63]. At this loading, $\tau \approx$ 3-13, compared with 0.4-0.7 in the measurements, so Eq.(3) gives the attenuation of the coherent field, an upper bound on the loss observed by a receiver that also collects forward-scattered power [60, 61].

In terms of the size parameter $x = \pi D_c f/c$, all spectra share the structure of the Lorenz-Mie extinction efficiency of a moderately refracting, absorbing sphere [58, 67]. They show a pronounced maximum at $x \approx$ 2.1-2.4 (near 88 and 70 GHz for $D_c$ = 2.55 and 3.108 mm), shoulders, a shallow minimum and a second maximum, and a plateau on which the loss scales with the geometric cross section. This envelope arises from the interference between the waves transmitted through and diffracted around the kernel, governed by the phase-shift parameter

$$\rho = 2x(n_r - 1) = \frac{2\pi D_c f}{c}(n_r - 1) \tag{6}$$

where $n_r$ is the real part of the refractive index $n_r + i\kappa = \sqrt{\varepsilon_{\text{eff}}}$ .

In the anomalous-diffraction approximation, the extinction maxima occur near $\rho \approx$ 4.1 and 10.8 and are spaced by $\Delta f = c/[D_c(n_r - 1)] \approx$ 121-138 GHz for $D_c$ = 3.108 mm and $n_r \approx$ 1.7-1.8 [67, 68], consistent with the maxima near 60 and 188 GHz in Fig. 3(b). The finer ripple of partial-wave resonances is largely damped by absorption and survives only as weak shoulders [69, 70]. Unlike high-index, low-loss spheres [71, 72], moist wheat combines a low refractive index with substantial absorption ($\kappa \approx$ 0.1-0.3), so its lowest multipoles overlap and Fig. 3 shows an interference structure rather than isolated Mie resonances; the magnetic-dipole condition $\lambda \approx n_r D_c$ of [71] ($x \approx 1.7$) lies below the first maximum.

Moisture modifies this structure in two ways. First, raising nr and, through swelling, Dc shifts all features to lower frequencies; for $D_c$ = 3.108 mm, the first maximum moves from about 70 to 60 GHz between 8% and 14% MC. Second, raising κ attenuates the wave transmitted through the kernel and weakens its interference with the diffracted wave [58, 73], so the extrema become shallower and the first minimum disappears at 14% MC. The curves therefore cross: wetter kernels attenuate more on the rising side of an extinction maximum, whereas on the falling side the scattering decreases with moisture faster than the absorption increases, so drier kernels can attenuate more, i.e., at 94 GHz for $D_c$ = 3.108 mm and between the sensitivity nulls near 90 and 155 GHz for $D_c$ = 2.55 mm [Fig. 3(d)]. At higher frequencies, the kernel becomes opaque: the transmitted wave is absorbed, the interference vanishes, and the extinction tends to twice the geometric cross section irrespective of composition (extinction paradox) [58, 74]. Despite the persisting dielectric contrast, the moisture dependence then becomes geometric: swelling raises the loss by about 1% per 1% MC for $x > 4$, without which the change would oscillate about zero. Most measurements of Fig. 2 ($x \approx$ 3.9-10.4) lie in this regime, which explains why the measured loss follows the kernel count.

### B. Frequency Selection

To quantify these effects, we define the moisture sensitivity and its relative form as,

$$S(f) = \frac{\partial L}{\partial M}, \qquad R(f) = \frac{1}{L}\frac{\partial L}{\partial M} \tag{7}$$

where $L = L_{\text{wheat}} = \alpha L_{\text{fg}}$ is the power loss in dB. $S$ (in dB per 1% MC) scales with the number of kernels along the path, whereas $R$ depends only on the single-kernel extinction. Both are evaluated at a fixed number of kernels and include the swelling. A calibration must also include the moisture dependence of the kernel count at a given mass flow, a geometric term of comparable magnitude [75], and the uncertainty of the swelling law. As Fig. 3(d) shows, $|S|$ forms lobes separated by minima near the crossings of the loss curves.

The dielectric contribution to $S$ alternates in sign between adjacent lobes, because the shift and damping of the interference structure raise the loss on one flank and lower it on the other. The swelling adds a positive contribution of about 1% of the loss per 1% MC, which dominates once the dielectric oscillation has weakened. For $D_c$ = 3.108 mm, $S$ therefore changes sign only near 68 and 126 GHz and remains positive up to 400 GHz.

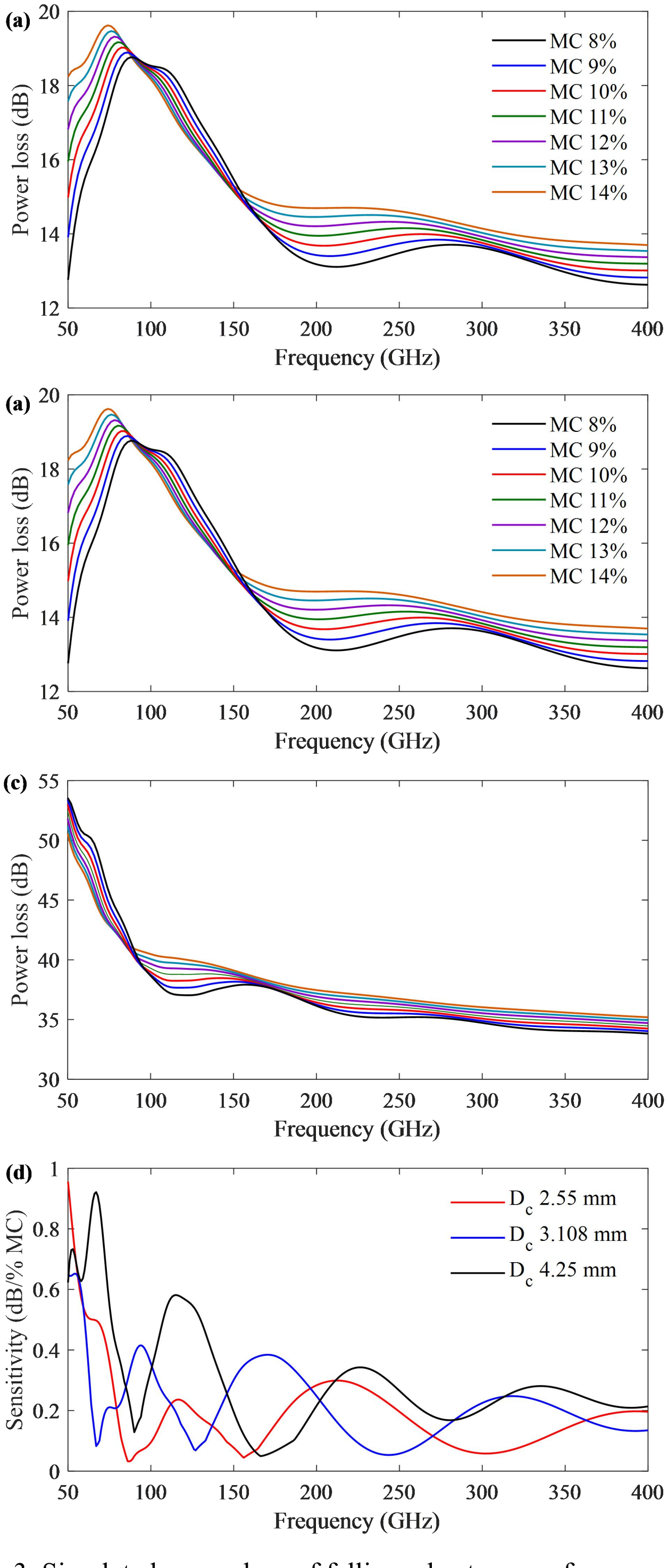

Fig. 3. Simulated power loss of falling wheat versus frequency at characteristic wheat diameters of (a) 2.55 mm, (b) 3.108 mm, and (c) 4.25 mm. (d) Moisture sensitivity versus frequency for the three characteristic wheat diameters. Panels (b) and (c) share the legend of (a).

The first lobe above the extinction maximum, on its falling flank where the loss decreases with moisture, carries the strongest dielectric response. Its peak lies near 116, 94, and 66 GHz for $D_c$ = 2.55, 3.108, and 4.25 mm, i.e., at $\lambda \approx D_c$ ($x \approx 3$), as the scale invariance of Lorenz–Mie theory requires [58]. This is not a resonance but the point at which the extinction efficiency responds most strongly to the complex refractive index, both mechanisms of Section III-A acting together. More than half of this response stems from the absorption index, the less constrained part of the dielectric model, so the position of the lobe is more robust than its magnitude. The swelling term weakens this negative lobe and strengthens the following positive one, which becomes comparable for $D_c$ = 3.108 mm and larger for 2.55 mm. At a fixed number density, the first-lobe peak grows roughly as $D_c^3$, whereas the subsequent lobes approach the swelling level of about 1% per 1% MC.

### *C. Channel Length and Loading*

To examine the influence of the channel propagation path, we varied the channel length through falling wheat for $D_c$ = 3.108 mm at 94, 171, and 317 GHz, three consecutive sensitivity maxima in Fig. 3(d), with $N = 1.6\times10^6$ kernels/m³ in a 0.5 m×0.5 m curtain. The loss grows linearly with the channel length (Fig. 4), the Beer-Lambert behavior implied by Eq.(3), and between 8% and 14% MC the specific attenuation decreases monotonically at 94 GHz but increases at 171 and 317 GHz (Table I). Only at 94 GHz does the dielectric response dominate; at 171 and 317 GHz, more than half of the sensitivity stems from the assumed swelling.

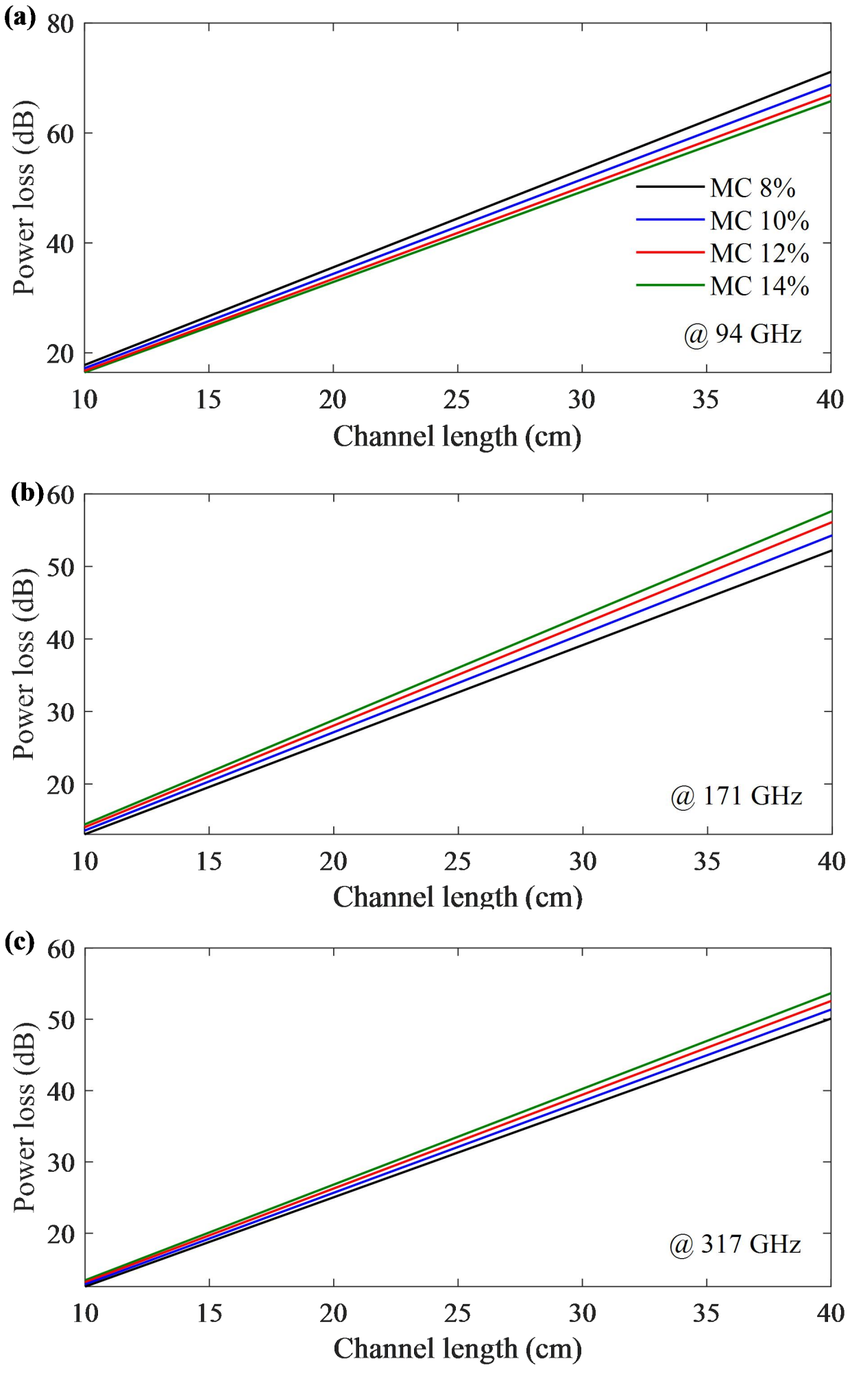

Fig. 4. Simulated power loss versus channel length in falling wheat for a characteristic kernel diameter of 3.108 mm at (a) 94 GHz, (b) 171 GHz, and (c) 317 GHz. The falling-wheat curtain has a cross-sectional area of 0.5 m × 0.5 m. Panels (b) and (c) share the legend of (a).

TABLE I
CHANNEL LENGTH IN FALLING WHEAT FOR 1 DB POWER LOSS SEPARATION UNDER 2% MOISTURE CONTENT CHANGE

| Frequency (GHz) | MC pair | Channel length for 1-dB difference (cm) | Power loss at this length (dB) |
|---|---|---|---|
| 94 | 8%-10% | 13.64 | 24.79-23.79 |
| 94 | 10%-12% | 17.55 | 30.61-29.61 |
| 94 | 12%-14% | 27.86 | 47.01-46.01 |
| 171 | 8%-10% | 16.53 | 21.31-22.31 |
| 171 | 10%-12% | 19.36 | 26.12-27.12 |
| 171 | 12%-14% | 24.03 | 33.66-34.66 |
| 317 | 8%-10% | 27.15 | 33.77-34.77 |
| 317 | 10%-12% | 29.94 | 38.34-39.34 |
| 317 | 12%-14% | 34.67 | 45.56-46.56 |

Table I gives the channel length needed for a 1-dB loss difference per 2% MC: about 14-35 cm at the simulated density, inversely proportional to it. Independently of the loading, a 2% MC step changes the loss on average by only 2.4%-3.5%, i.e., by 0.05-0.11 dB at the measured losses of Section II.. Moreover, since the loss is proportional to the loading, a 1% error in the number of kernels along the path mimics 0.5%-0.8% MC, whereas the loading varied by up to 26% between runs, which calls for a loading-independent observable. Because Eq. (3) is linear in the number density, the ratio of the losses at two frequencies $f_a$ and $f_b$,

$$\Gamma(f_a, f_b; M) = \frac{L(f_a, M)}{L(f_b, M)} = \frac{\int N(r)\sigma(r; f_a, M)\,\mathrm{d}r}{\int N(r)\sigma(r; f_b, M)\,\mathrm{d}r} \tag{8}$$

is independent of the number density and the channel length and depends only on the moisture, the temperature, and the size distribution, the scattering-regime counterpart of the density-independent functions of microwave moisture sensing [76, 77]. Adjacent lobes of opposite sign are most effective: Γ(94 GHz, 171 GHz) falls from 1.41 at 8% MC to 1.15 at 14% MC (-3.1% per 1% MC), more than either frequency alone. Alternatively, the lobe at $x \approx 3$ can be paired with the opaque plateau, where the loss depends mainly on the loading and the kernel cross section. Kernel size and temperature must still be known or compensated for: a 10-K temperature change [45] or a 2.5% diameter change shifts Γ by about 3.5% or 2%, each equivalent to about 1.1% or 0.6% MC.

*D. Received-Power Fluctuations*

The falling wheat also causes temporal fluctuations of the received power, quantified by an effective Rician $K$ factor, the power ratio of the dominant LoS component to the diffuse scattered components [78]. For each frequency-moisture condition, maximum-likelihood Rician fits to the readings of the stable falling interval (69-107 per record, nine records) give $K=s^2/(2\sigma^2)$, which was averaged over the records. At the beam center, the kernels move at about 2.45 m/s (free fall over 28.8 cm from 0.6 m/s) and cross the 11.07-cm beam in about 45 ms, so the 5 Hz readings, about 4.4 transit times apart, are independent realizations of the kernel arrangement, as required for estimating the mean loss and the fluctuation statistics.

Fig. 5 shows the empirical cumulative distribution functions (CDFs) of the received signal-to-noise ratio (SNR) at 140, 220, and 320 GHz. The Rician fits represent the data well. Averaged over frequency, $K$ is 22.05, 22.55, and 19.52 dB for 7.8%, 11%, and 14% MC, representing a received-power standard deviation of about 0.5-0.65 dB [79, 80], two orders of magnitude above the free-space stability. These wheat-induced fluctuations leave the channel strongly LoS dominated and follow the loading rather than the moisture. Independent, Poisson-distributed kernels with mean number $N_b$ would make the loss fluctuate by about $L/\sqrt{N_b}$ , i.e., $K \approx 2N_b/\tau^2$, so the measured $K$ corresponds to only about 20 effective kernels, an order of magnitude fewer than the approximately 250 in the beam. The fluctuations therefore stem from collective variations of the flow, such as kernel clustering and feeder pulsation, which modulate the loading as a whole: loading-proportional fluctuations ($K \propto 1/\tau^2$) predict about 2.5 dB between the 11% and 14% samples, close to the 3 dB observed, whereas independent kernels ($K \propto 1/\tau$) predict only about 1 dB. The Racian factor $K$ thus measures a flow-induced modulation of the extinction rather than multipath fading. The corresponding optical-depth fluctuation, $\sigma_\tau \approx (2/K)^{1/2} \approx 0.10$-$0.15$ is more than six times that implied by the $K \approx 38$ dB reported for fog, rain, and snow with comparable setups [42, 81, 82], reflecting the unsteady loading of a mechanically fed flow. Being common to all frequencies, this modulation cancels in the two-frequency ratio of Eq.(8) when both frequencies probe the same volume simultaneously.

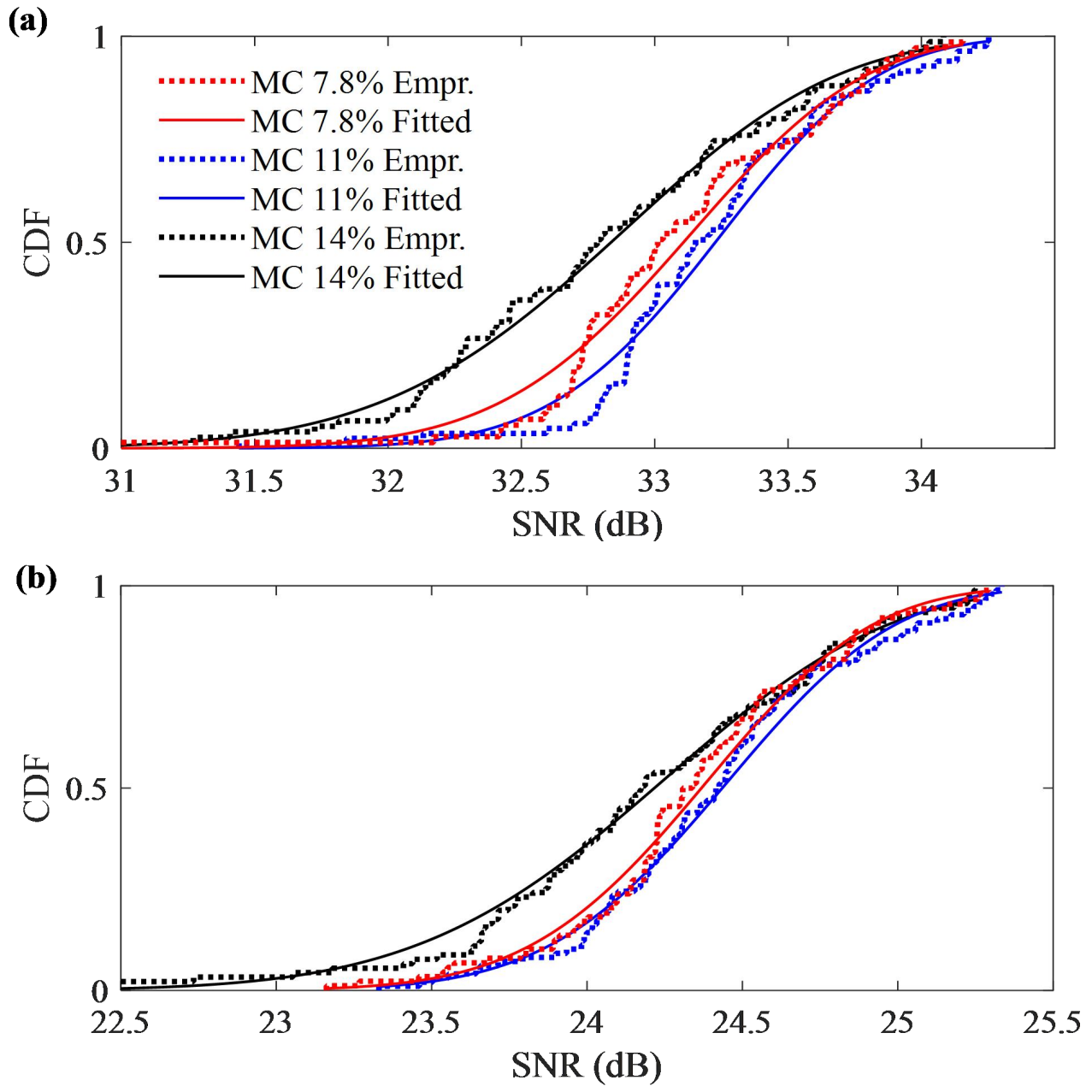

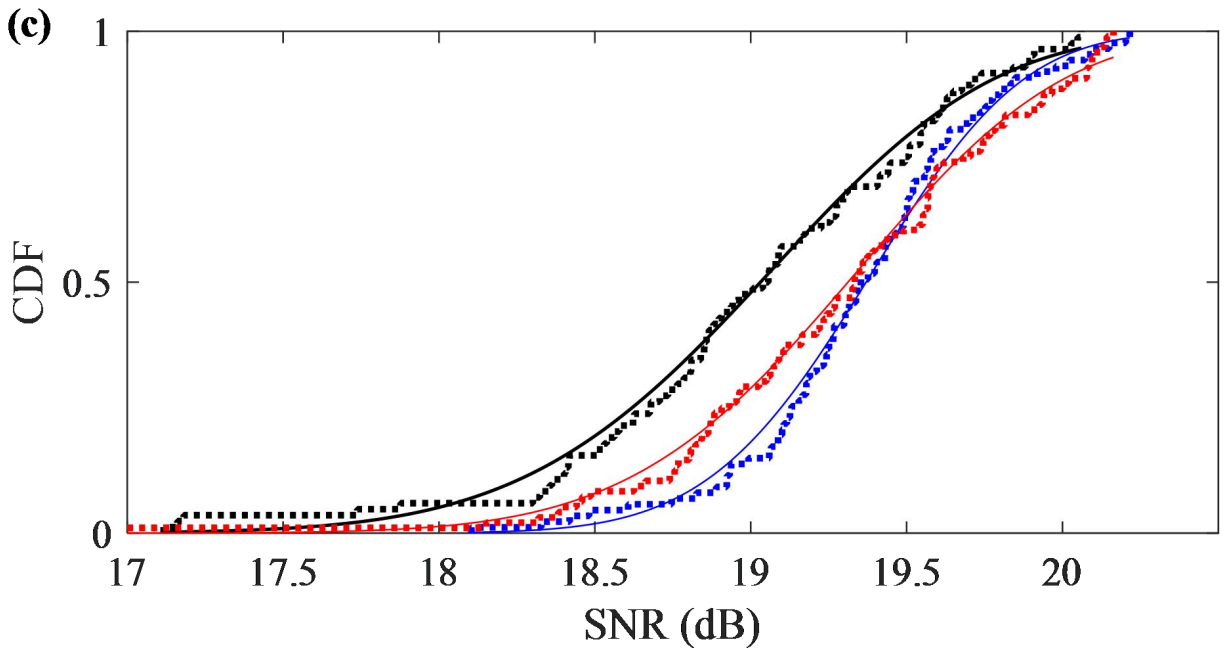

Fig. 5. Cumulative distribution function (CDF) of the received SNR of the (a) 140-GHz, (b) 220-GHz, and (c) 320-GHz channels in falling wheat, fitted with the Rician model. Panels (b) and (c) share the legend of (a).

## IV. Bit Error Rate Performance

A sensing channel through falling wheat is also a communication channel, and in an integrated sensing and communication system the same waveform would serve both purposes [23, 24], as in rainfall estimation from microwave-channel attenuation [83]. Because the channel is LoS dominated, the average SNR follows from a deterministic channel budget that includes the loss induced by the falling wheat,

$$\mathrm{SNR}(f, M, D_c, L_{fg}) = P_t + G_t + G_r - L_{FSPL}(f, d) - L_{wheat}(f, M, D_c, L_{fg}) - P_N \quad (9)$$

where $P_t$ is the transmit power and $G_t = G_r = 33$ dBi are the antenna gains measured at 220 GHz and taken as frequency independent. $L_{FSPL}(f,d)$ is the free-space path loss over a channel distance $d = 10$ m, calculated according to Recommendation ITU-R P.525-5 [84]. $L_{wheat} = \alpha L_{fg}$ is the loss calculated from Eq. (3) for $L_{fg} = 0.15$ m and $N = 1.6\times10^6$ m$^{-3}$, the case of Fig. 3, which gives 24.69 dB at 100 GHz for $D_c = 3.108$ mm and 11% MC. $P_N = -50$ dBm is the receiver noise power. QPSK and 16-QAM are considered because both have been employed in high-speed low-THz and THz transmission demonstrations [84, 85] and are specified for 252-322 GHz channels in IEEE Std 802.15.3d [86].

For an instantaneous SNR obtained from Eq.(9), equal to the symbol energy-to-noise ratio $E_s/N_0$ when the noise bandwidth equals the symbol rate, the conditional BERs are

$$\mathrm{BER_{QPSK}} = \frac{1}{2}\mathrm{erfc}\left(\sqrt{\gamma}\right) \tag{10}$$

$$\mathrm{BER_{16QAM}} = \frac{3}{8}\mathrm{erfc}\left(\sqrt{\frac{2\gamma}{5}}\right) \tag{11}$$

which are the standard expressions for Gray-coded QPSK and 16-QAM in additive white Gaussian noise (AWGN) [87, 88]. The latter in its usual nearest-neighbor form, and are also adopted in terahertz channel-level BER evaluations [89]. For the Rician channel, the average BER is obtained over the instantaneous SNR distribution [42, 90], as

$$\overline{\mathrm{BER}} = \int_0^{\infty} \mathrm{BER}(\gamma) f(\gamma)\,\mathrm{d}\gamma \tag{12}$$

$$f(\gamma) = \frac{1+K}{\bar{\gamma}}\exp\left(-K - \frac{(1+K)\gamma}{\bar{\gamma}}\right) I_0\left(2\sqrt{\frac{K(1+K)\gamma}{\bar{\gamma}}}\right) \tag{13}$$

where $\bar{\gamma}$ is the average *SNR* obtained from Eq. (9), $K$ is the Rician factor, and $I_0$ is the zeroth-order modified Bessel function of the first kind. Because a constant $K$ of 22 dB, representative of the measured range, was used for all moisture conditions, Eq. (12) represents the long-term average over the flow-induced fluctuations of Section III-D. At the simulated loading, this value is also close to the independent-kernel value $K \approx 2N_b/\tau^2$ of Section III-D, which gives about 20-24 dB for $D_c$ = 3.108 mm ($N_b \approx 2300$ kernels in an 11 cm beam, $\tau \approx 4.2$-6.8). Fig. 6 therefore represents a steady flow, whereas collective variations such as those of the feeder used here would lower $K$. Equation (12) was evaluated by Monte Carlo sampling [91] and checked by numerical quadrature of its moment-generating-function form [90].

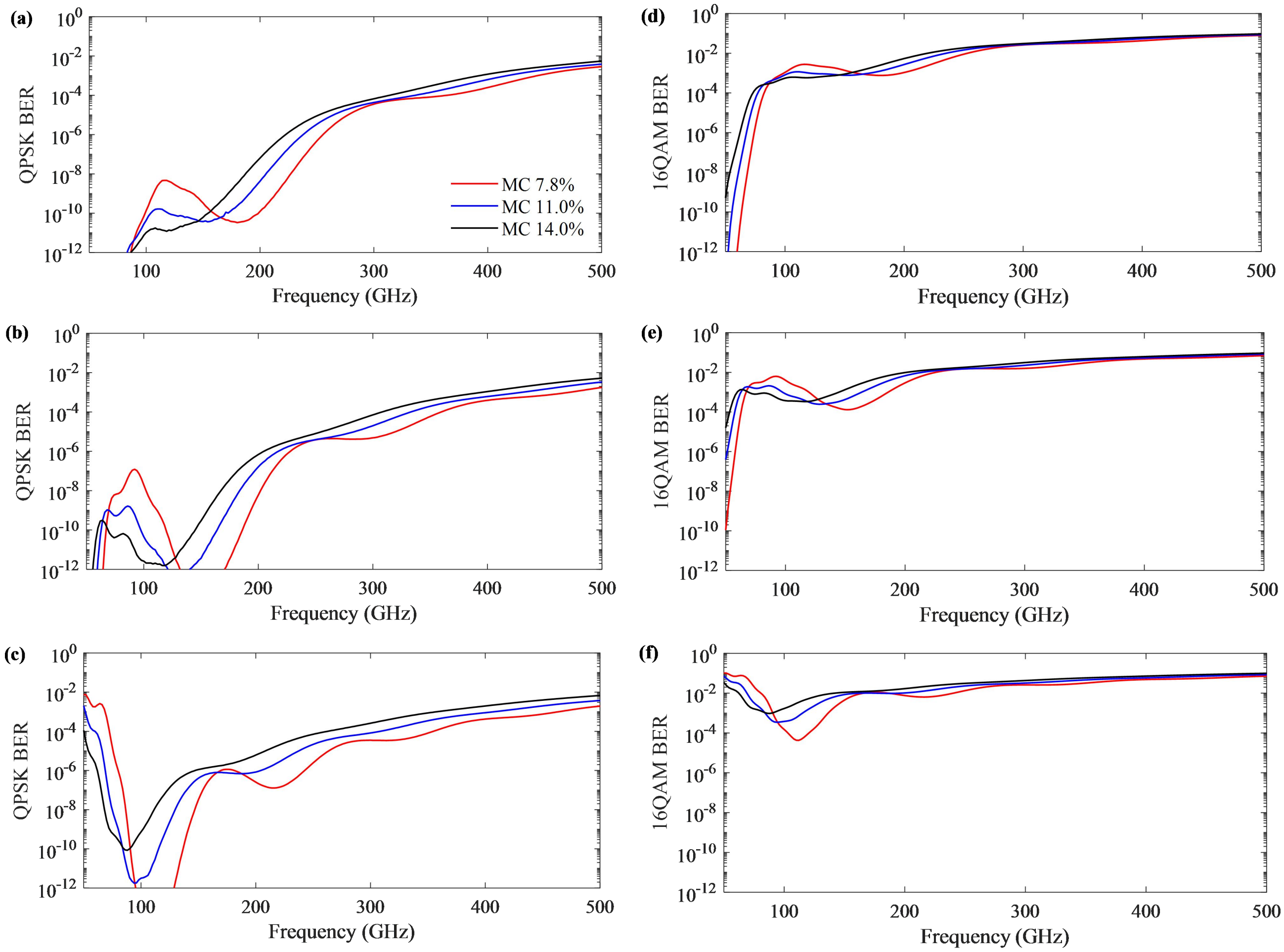

Fig. 6. BER performance versus frequency for different moisture contents in falling wheat. QPSK BER at characteristic wheat diameters of (a) 2.55 mm, (b) 3.108 mm, and (c) 4.25 mm, and 16-QAM BER at characteristic wheat diameters of (d) 2.55 mm, (e) 3.108 mm, and (f) 4.25 mm. $P_t$ is set to 12, 18, and 33 dBm for the characteristic diameters of 2.55, 3.108, and 4.25 mm, respectively. Panels (b)-(f) share the same legend of (a).

Fig. 6 shows the predicted BER of QPSK and 16-QAM. Its frequency dependence reflects the competition between the two frequency-dependent terms of Eq.(9) - the Mie extinction, which decreases beyond its first maximum, and the free-space path loss, which increases as 20log($f$). The resulting interior BER minimum moves to lower frequencies as the kernels become larger, so the operating frequency must be selected jointly with the kernel size. The 16-QAM curves show the same features at much higher BERs, more than seven orders of magnitude above QPSK at 100 GHz for $D_c$ = 3.108 mm.

Because the curves of a panel share the same transmit power and path loss, their separation reflects only the moisture-dependent loss differences of Fig. 3. At 126 GHz, for example, the loss differs by only 0.51 dB across the three moisture contents for $D_c$ = 3.108 mm, near its sensitivity null, but by 3.41 dB for $D_c$ = 4.25 mm, which separates the QPSK curves for 7.8% and 14% MC by about 5-6 orders of magnitude. Above about 300 GHz, the curves draw together because the larger path loss moves the operating point to BERs of $10^{-5}$-$10^{-2}$, where the BER characteristic is less steep. In the waterfall region, the BER changes by about 0.5-1.5 decades per dB at BERs of $10^{-3}$–$10^{-9}$ ($K$ = 22 dB), so a 1-dB loss increase, the criterion of Table I for a 2% MC step, raises the BER by a factor of about 3-30. A decibel-level loss change thus appears as an order-of-magnitude change of a quantity that a digital receiver provides as a byproduct, making the BER a far more responsive moisture indicator than the received power.

## V. Conclusion

Safe wheat storage calls for online moisture monitoring at grain intake, where the grain is in motion and must be measured without contact, yet THz waves, though highly sensitive to water, have so far been applied only to static grain samples. This article therefore investigated moisture sensing through a low-THz channel across falling wheat. The channel power loss was measured at six frequencies from 120 to 320 GHz for three moisture contents, interpreted with a physics-based model that couples the double-Debye permittivity of water, Landau-Lifshitz-Looyenga mixing, and Lorenz-Mie extinction, and translated into the BER of QPSK and 16-QAM channels.

The model agrees with the measured loss, confirming that it captures the THz power loss in falling wheat to first order, and the measured Rician K factors of 20-23 dB show that the channel remains line-of-sight dominated. Simulations further show that, above the extinction maximum, the dielectric moisture response is strongest where the wavelength approaches the kernel diameter, which provides a simple rule for choosing the operating frequency according to the kernel size. The ratio of the losses at two frequencies cancels the kernel loading and the channel length and thus serves as a loading-independent moisture observable. Most importantly, channel-level analysis shows that the BER converts a 1 dB moisture-induced change in the channel loss into a 3- to 30-fold change, making it a highly responsive, communication-native moisture indicator that a digital receiver provides as a byproduct of data transmission. Future work will verify these predictions with modulated-channel measurements.